# Skyrmion Generation in a Plasmonic Nanoantenna through the Inverse Faraday Effect

Xingyu Yang[1], Ye Mou[1], Bruno Gallas[1], Sébastien Bidault[2], and Mathieu Mivelle[1,*]

[1]Sorbonne Université, CNRS, Institut des NanoSciences de Paris, INSP, F-75005 Paris, France

[2]Institut Langevin, ESPCI Paris, Université PSL, CNRS, F-75005 Paris, France

*Corresponding authors:

mathieu.mivelle@sorbonne-universite.fr

**Abstract**

Skyrmions are topological structures characterized by a winding vectorial configuration that provides a quantized topological charge. In magnetic materials, skyrmions are localized spin textures that exhibit unique stability and mobility properties, making them highly relevant to the burgeoning field of spintronics. In optics, these structures open new frontiers in manipulating and controlling light at the nanoscale. The convergence of optics and magnetics holds therefore immense potential for manipulating magnetic processes at ultrafast timescales. Here, we explore the possibility of generating skyrmionic topological structures within the magnetic field induced by the inverse Faraday effect in a plasmonic nanostructure. Our investigation reveals that a gold nanoring, featuring a dark mode, can generate counter-propagating photocurrents between its inner and outer segments, thereby enabling the magnetization of gold and supporting a skyrmionic vectorial distribution. We elucidate that these photocurrents arise from the localized control of light polarization, facilitating their counter-propagative motion. The generation of skyrmions through the inverse Faraday effect at the nanoscale presents a pathway towards directly integrating this topology into magnetic layers. This advancement holds promise for ultrafast timescales, offering direct applications in ultrafast data writing and processing.

# 1. Introduction

In recent years, the field of condensed matter physics has witnessed a surge of interest in a class of intriguing topological structures known as skyrmions.[1] These nanoscale whirlpools of magnetic order, originally conceived as solutions to equations of nuclear physics by Tony Skyrme in the early 1960s,[2] have emerged as promising candidates for revolutionizing information storage and processing technologies.[3] In magnetic materials, skyrmions are localized spin textures that exhibit unique stability and mobility properties, making them highly relevant to the burgeoning field of spintronics.[4] The distinctive feature of skyrmions lies in their nontrivial topology, characterized by a winding configuration of spins that provides a quantized topological charge to these entities.[5] Unlike traditional magnetic domains, skyrmions possess a particle-like nature and can exist as individual entities or form periodic arrays in magnetic materials.[6] Their stability at nanoscale dimensions and the ability to manipulate them with low-energy currents have positioned skyrmions as potential building blocks for next-generation magnetic memories and logic devices.[7]

The allure of skyrmions extends beyond the realm of condensed matter physics, permeating diverse fields of research with their unique topological characteristics and dynamic behaviors. They have, for instance, manifested in unexpected domains such as optics,[8-10] opening new frontiers in the manipulation and control of light at the nanoscale. In fact, the unique topological features of skyrmions, characterized by their swirling vectorial configurations, introduce intriguing possibilities for tailoring optical properties and creating innovative devices in photonics. Moreover, in recent years, researchers have explored the interaction between skyrmions and light, uncovering fascinating phenomena that bridge the fields of magnetism and optics.[11] The integration of skyrmions into photonic structures also holds promise for developing all-optical information processing devices,[12] and quantum optics.[13] Skyrmions, with their nanoscale dimensions and stability, can serve as dynamic elements in optical circuits, enabling the transmission and manipulation of information through the controlled motion of these magnetic textures. This burgeoning field of research not only extends the functionalities of existing photonic technologies but also introduces novel paradigms for the design of next-generation optical devices.

Here, we propose to go one step further in the interaction between light and magnetism at the nanoscale by developing a model of plasmonic nanostructures enabling the generation of a magnetic field by inverse Faraday effect (IFE) bearing the topological structure of a skyrmion. The IFE is a magneto-optical process enabling the magnetization of matter by optical excitation. While known since the 1960s,[14-16] this physical phenomenon has recently garnered attention due to advancements in nanophotonics and ultrafast optics.[17-26] The manipulation of light at the nanoscale, specifically with regard to its polarization, gradients, and amplitude, has been demonstrated to yield ultrafast and intense magnetic fields at this scale.[19,21] These properties offer avenues for manipulating magnetic processes at ultrafast timescales and nanoscopic spatial dimensions. In this paper, we establish that a ring-shaped plasmonic nanostructure generates a skyrmionic distribution of magnetic fields through IFE. This novel behavior arises from the production of two drift countercurrents within the metal of the ring, notably in its inner and outer segments. These opposite drift currents result from the manipulation of the polarization of light around the plasmonic antenna, achieved through the specific excitation of the anti-bonding mode of the plasmonic nanoantenna.

The significance of these findings lies in their potential to facilitate the implementation of skyrmionic topological structures within a magnetic layer through direct magnetic action and potentially at ultrafast timescales. This breakthrough, therefore, opens avenues for developing next-generation magnetic memory and logic devices.

## 2. Results:

A skyrmion can exhibit diverse topological structures.[5] However, our focus in this article centers exclusively on the Néel type. Illustrated in Figures 1a and b, we outline the vector distribution of the magnetic field within this specific manifestation of a skyrmionic structure. This type is characterized by a continuous rotation of the magnetic field in the plane of the skyrmion, extending seamlessly from its center to its periphery, forming an intricate swirling pattern.[27] Figures 1c and d depict one method of reaching such a vectorial distribution in the magnetic field, involving the generation of two counter-propagating currents with different radii.

For the creation of this intricate vectorial distribution within a nanoscale magnetic field, we propose the use of a ring-shaped plasmonic nanoantenna, as illustrated in Figures 1e and f, generating counter-propagating drift currents through IFE, as illustrated in Figure 1f. To this end, the nano-ring is positioned on a glass substrate and excited by a right-circularly polarized plane wave incident from the substrate side (Figure 1e,f).

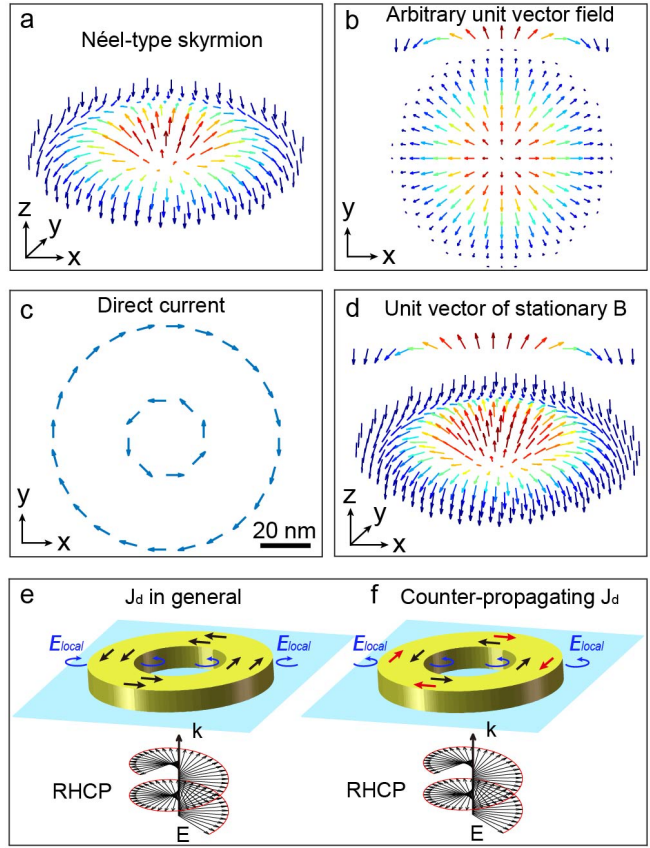

**Figure 1**. Description of a Néel-type skyrmion. a) 3D and b) normal perspective of unit vector distributions of a Néel-type skyrmion. The arrows represent the orientation of the vectors, and the colors are the amplitude of their Z component. c) Distribution of direct electric currents in an XY plane, required for the realization of d) a Neel-type magnetic skyrmion distribution for a rotation of the vector in the XZ and YZ planes. e and f) Schematic description of the plasmonic nanostructure considered in this study to generate a skyrmionic distribution by IFE: a gold nanoring is placed on a glass substrate and excited by a right circularly polarized plane wave incident from the substrate side. Two cases are considered: drift currents ($\bm{J_d}$) generated by IFE propagate e) in the same direction and f) in opposite directions between the inner and outer parts of the nanoring. The blue arrows represent the local polarization required to generate the drift currents shown in the drawing, black and red arrows indicate the direction of drift currents.

The theoretical description of the drift currents ($\bm{J_d}$) induced by IFE in a metal has been extensively documented.[28,29] The ensuing Equation describes this phenomenon:

$$\bm{J_d} = \frac{1}{2en} Re\left(\left(-\frac{\nabla \cdot (\sigma_\omega \bm{E})}{i\omega}\right) \cdot (\sigma_\omega \bm{E})^*\right) \qquad (1)$$

Here e represents the charge of the electron (e < 0), n is the charge density at rest, $\sigma_\omega$ denotes the dynamic conductivity of the metal, and **E** corresponds to the optical electric field.

Leveraging these currents and employing the Biot and Savart equation (Equation 2), the vectorial distribution of the magnetic field can be computed in the space surrounding the nanoantenna.

$$\boldsymbol{B} = \frac{\mu_0}{4\pi} \iiint \frac{\boldsymbol{J_d} \times \boldsymbol{r}}{|\boldsymbol{r}|^3} dV \qquad (2)$$

Equation 1 implies that the prerequisite for inducing counter-propagating currents on opposing sides of the nanoring is the presence of light, both inside and outside, carrying the same type of polarization helicity, as depicted in Figure 1f.

Figure 2a illustrates the spectral response of the electric field enhancement at the XYZ center of a nanoring, featuring an inner diameter "d" of 120 nm and an outer diameter "D" of 240 nm with thickness 30 nm. Notably, for these dimensions, the structure exhibits two resonances in its spectral response, with one being weaker around 570 nm and the other more pronounced at approximately 1100 nm. Let's first delve into the characteristics of the stronger resonance. In Figure 2b, the electric field distribution in an XY plane at the Z center of the structure is outlined at a wavelength λ of 1100 nm. The field distribution indicates a notable increase in the electric field both inside and outside the nanoring at this specific wavelength.

Moving forward, Figure 2c shows the spin density distribution in the same plane as the field in Figure 2b. The spin density, as defined by Equation 3, represents a vectorial physical quantity that illustrates the polarization state of light within a given plane. Figure 2c displays the Z component of the spin density describing the helicity of light in the XY plane. This density can reach positive or negative values, corresponding to right or left elliptical polarizations. Specifically, a positive spin density in our reference system implies a right helicity, a negative spin density corresponds to a left helicity, and a zero density denotes linear polarization. In the far field, the spin density is constrained between -1 and 1, with -1 indicating left circular polarization and 1 denoting right circular polarization. However, in the near field, when normalized by the incident intensity $|E_0|^2$, the spin density can reach significantly larger values due to the enhanced fields, giving rise to the concept of super-circular light,[25] drawing an analogy with super-chiral light.[30]

$$s = \frac{1}{|E_0|^2} Im(E^* \times E) \qquad (3)$$

From Figure 2c, it is clear that the spin densities within and outside the ring exhibit opposite signs. This local light polarization state does not provide the prerequisite for counter-propagating drift currents, as explained earlier and detailed in Figure 1e,f. Consequently, in this scenario, the currents will flow in the same direction, as illustrated in Figure 2d, and they will not generate a skyrmionic structure through IFE.

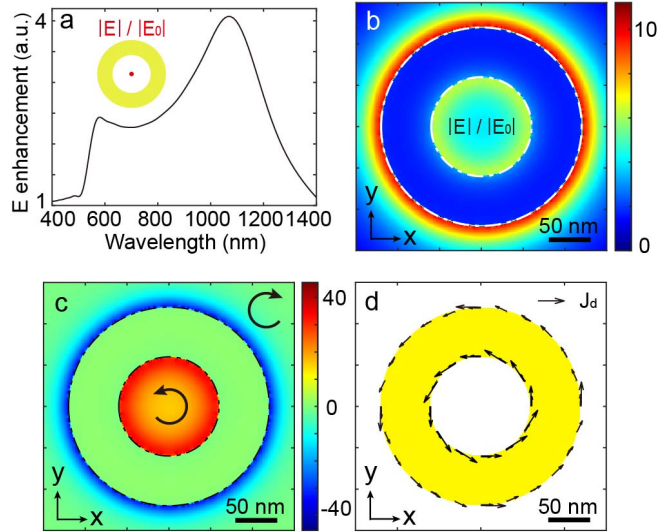

**Figure 2**. Description of the physical behavior of the nanoring. a) Spectral response in terms of electric field enhancement at the XYZ center of a nanoring with an inner diameter of 120 nm, outer diameter of 240 nm, and thickness of 30 nm. Spatial distribution in an XY plane at the Z center of the nanoantenna for b) electric field enhancement normalized by the incident wave, c) spin density, and d) drift currents.

Nevertheless, these annular plasmonic structures are known in the literature for supporting a diversity of modes.[31-33] Specifically, a nanoring can be seen as a combination of two distinct structures: a circular nano-aperture and a nanodisk, each exhibiting a resonance mode (Figure 3a and Figure S1). The combination of these structures results in a nanoring with the ability to sustain two distinctive coupled modes (Figure 3a): a bonding mode (commonly referred to as bright) and an anti-bonding mode (commonly referred to as dark). The manifestation of these modes becomes clear when studying the spectral response of the nanoantenna with respect to the charge density (Figure 3b) within the inner and outer metallic components of the antenna. Charge density distributions corresponding to the bonding and anti-bonding modes identified in Figure 3b are presented in Figures 3c and d, respectively. As anticipated, the charge densities between the inner and outer regions of the antenna exhibit an out-of-phase relationship for the anti-bonding mode, while they are in phase for the bonding mode.

Upon inspecting the spectral response regarding spin densities for the inner and outer segments of the nanoring (Figure 3e), it is evident that polarizations are opposite in the case of the bonding mode (at λ = 1100 nm in Figure 3e). In contrast, for the anti-bonding mode, the spin densities exhibit the same sign within a specific wavelength range (at λ = 570 nm in Figure 3e).

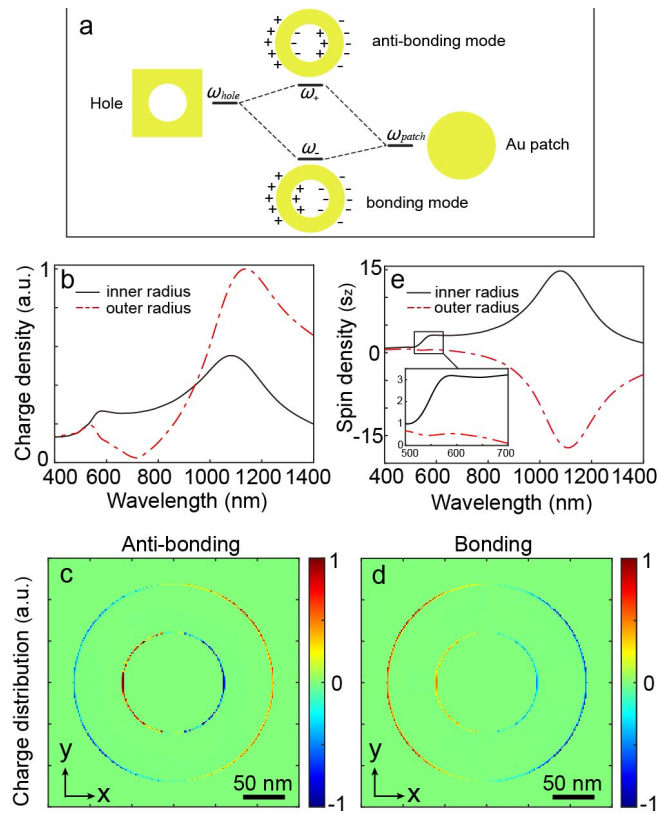

**Figure 3**. Multipolar behavior of the nanoring. a) A nanoring can be seen as the combination of two distinct structures, namely a nanohole and a nanopatch. Each of these structures has its resonance mode, and their coupling results in a structure itself carrying two resonance modes, one bright, known as bonding, at a lower frequency, and the other dark, known as anti-bonding, at a higher frequency. b) Spectral response in terms of electron density for the inner (black line) and outer (red dashed lines) parts of the metal. Spatial distributions of electron densities in an XY plane at the Z center of the nanoring for c) the anti-bonding mode and d) the bonding mode. e) Spectral response in terms of spin density for the inner (black line) and outer (red dashed line) parts of the nanoring. A zoom of the portion corresponding to the anti-bonding mode is shown in the inset.

To illustrate this difference, Figures 4a and b present spin density distributions in an XY plane at the Z center of the nanoring for the two modes inherent to this antenna. The contrasting polarizations are clearly visible in these depictions. Subsequently, Figures 4c and d exhibit the drift currents corresponding to the spin densities in Figures 4a and b, calculated using Equation 1. In the bonding mode, we observe that the drift currents propagate in the same directions within the inner and outer regions of the nano-antenna. In contrast, in the case of the anti-bonding mode, the currents exhibit counter-propagation on each side of the nanoring.

Consequently, this dark mode is anticipated to facilitate the generation of a skyrmionic topological distribution through IFE, as illustrated in Figures 1c and d. The dissimilarity in local polarization between these two modes is found in the dipolar nature of the bonding mode and the coupling between two dipolar modes within the anti-bonding mode. Comprehensive information on these divergent effects is provided in the supplementary information.

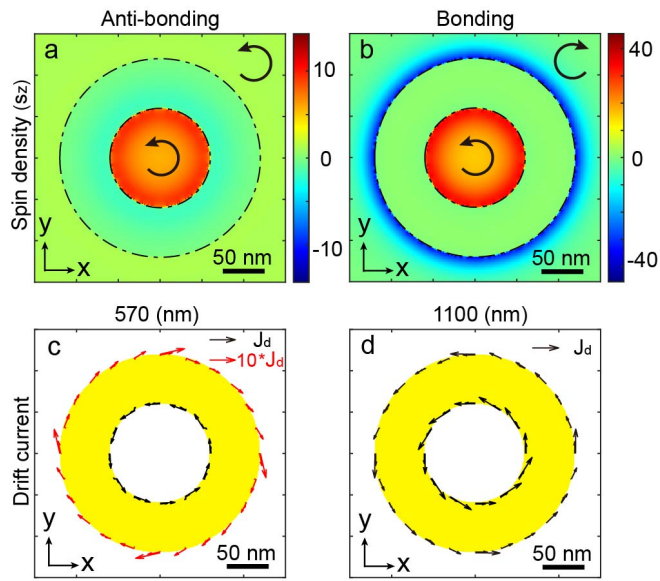

**Figure 4**. Comparison of dark and bright modes in the nanoring. Spatial distributions of spin densities in an XY plane at the Z center of the plasmonic nanostructure for a) the anti-bonding mode ($\lambda$ = 570nm) and b) the bonding mode ($\lambda$ = 1100 nm). Arrows represent the helicity of light. Spatial distributions of drift currents in an XY plane at the Z center of the nanoring for c) the anti-bonding mode and d) the bonding mode. The length of the arrows represents the relative amplitude of the generated photocurrents.

Having identified the physical process responsible for the generation of counter-propagating drift currents, an opportunity arises to investigate the parameters governing this phenomenon. Our focus centers on determining a configuration with dimensions capable of inducing drift currents of comparable magnitudes in both the internal and external regions of the nanoring, resulting in a uniform magnetic field amplitude. Figure 5 explores the set of parameters "d" and "D" to achieve a skyrmionic vector distribution of the magnetic field. In Figure 5a, the skyrmion number of the nanoring is presented for excitation at a wavelength of 600 nm, considering various values of "d" and "D" with a gold thickness of 30 nm. The skyrmion number "Q", an integral topological invariant used to characterize the winding of spins in the magnetic texture, is defined as an integer and is associated with the topology of the spin configuration within the skyrmion. "Q" is defined by unit vector **u**, expressed as:

$$Q = \frac{1}{4\pi} \iint \boldsymbol{u} \bullet \left(\frac{\partial \boldsymbol{u}}{\partial x} \times \frac{\partial \boldsymbol{u}}{\partial y}\right) dxdy \qquad (4)$$

In our study, the vectorial distribution closest to a perfect Neel-type skyrmionic topology is observed when the skyrmion number, Q, approaches ±1. The sign of Q dictates the up or down orientation of the magnetic field at the center of the antenna. Figure 5a illustrates that numerous pairs of parameters, "d" and "D", can yield a value of Q close to 1, indicating that various physical parameters can be manipulated to achieve a skyrmion. Despite the magnetic field's vector distribution taking on a skyrmionic topology, our objective was to develop a plasmonic nanoantenna with a relatively uniform magnetic field amplitude in this vectorial distribution. To this end, Figure 5b presents the ratio of spin density amplitudes between the inner and outer segments of the nanoring for different diameter values. This ratio aligns well with the ratio of drift countercurrents between the inner and outer metal components, as depicted in Figure 5c. By leveraging these drift currents, we calculate the ratio between the maximum and minimum magnetic fields on either side of the nanoring walls, as illustrated in Figure 5d. This figure provides a range of parameters that enable the skyrmionic distribution to exhibit a relatively homogeneous magnetic field amplitude.

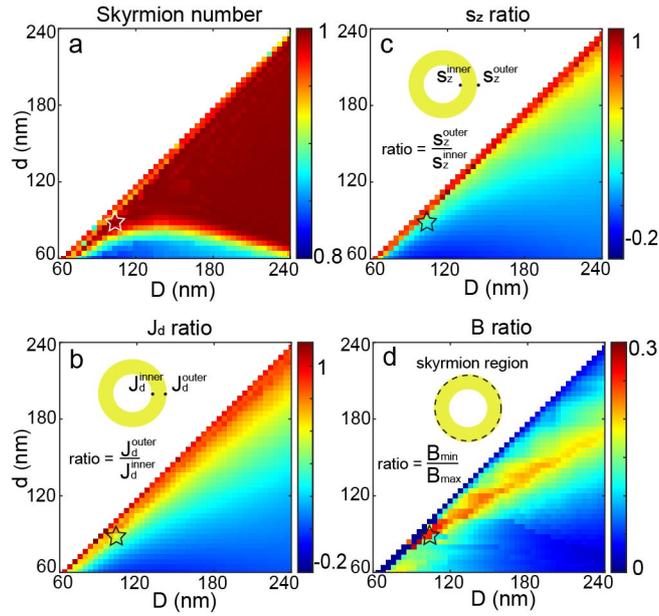

**Figure 5**. Study of skyrmion generation by IFE for different inner "d" and outer "D" nanoring diameters, with a 30 nm gold thickness and excitation at λ = 600 nm. a) Skyrmion number values. b) Ratio of spin density values along Z between the inner and outer parts of the plasmonic nanostructure. c) Ratio of drift currents (absolute values) associated with the spin densities shown in b) between the inner and outer parts of the nanoring. d) ratio between the maximum and minimum magnetic field amplitudes generated from the drift currents shown in c) within the skyrmionic region of the plasmonic nanoantenna. The star highlights the dimensions considered in Figure 6.

Building upon these findings, Figure 6 depicts the vectorial distribution of the magnetic field generated by IFE for a nanostructure derived from our simulations, considering two types of optical excitations: right-circularly polarized and left-circularly polarized. The selected nanoantenna is showcased in Figure 6a, having a thickness of 30 nm, an inner diameter of 88 nm, and an outer diameter of 104 nm. Figures 6b and c illustrate the distribution of the Z-component of the optically induced magnetic field, normalized by the magnetic field amplitude in the center, in case of right- and left-circular polarizations, respectively. Consistent with the IFE principle, these vector distributions are perfectly inverted in orientation, as shown in Figures 6d and e, presenting these fields in 3D at 1 nm from the gold surface of the antenna.

Figures 6d and e present the unit vector distributions of **B**. As observed, a skyrmionic topological distribution manifests on the surface of this antenna. The magnetic field is perpendicular to the surface at the center of the antenna, progressively rotating within the skyrmion plane from the center to the periphery of the antenna, forming a swirling pattern that completely reverses at the edge of the nanoantenna. To illustrate this vectorial motion in space, Figure 6f displays the orientation of the magnetic field and its unit vector for a linecut from the center to the edge of the antenna. The spatial inversion of the magnetic field is evident, with the magnetic field maintaining the same sign over the entire nanoantenna, and a flip of the magnetic field occurring at the end of the nanostructure.

This outcome is particularly noteworthy, aligning with the observed behavior in many skyrmionic spin structures found in magnetic materials.[34] These characteristics could facilitate the implementation of such vectorial structures in magnetic materials using plasmonic IFE. Additionally, the composition of magnetic materials defines the physical size of skyrmionic topologies. Thus, the ability to adjust the size of the vectorial distributions with this symmetry by changing the dimensions of the plasmonic nanostructures would be a valuable advantage for the all-optical generation of skyrmions.

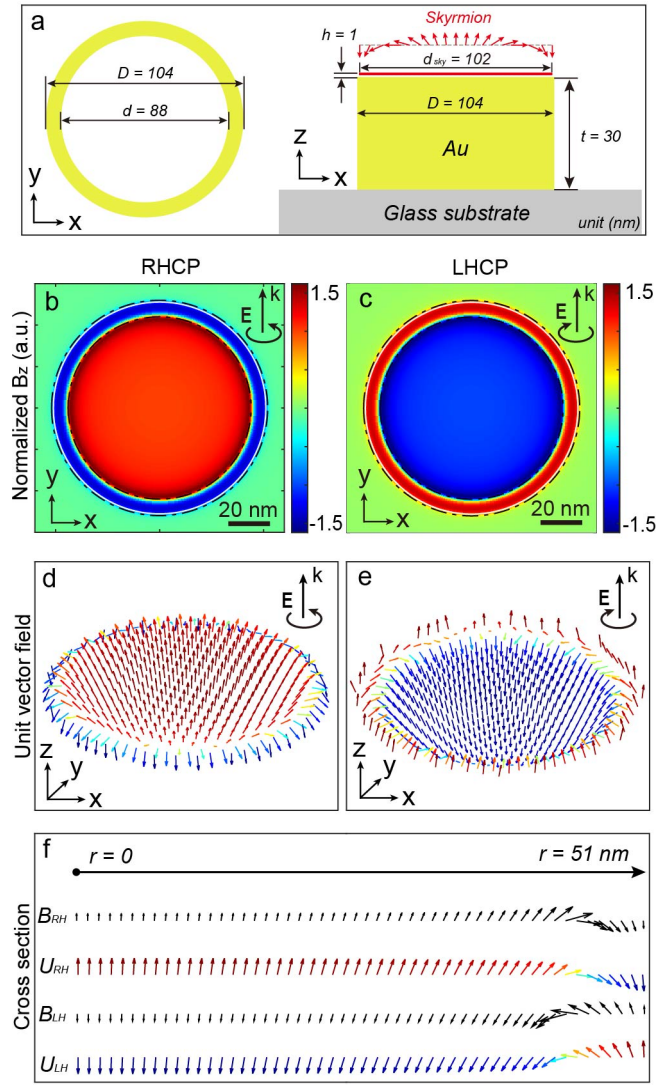

**Figure 6**. Example of a nanostructure generating a perfect Néel-type skyrmion for excitation at λ = 600 nm. a) Schematic representation of a nanoring enabling the generation of a Néel-type skyrmion by IFE. Its dimensions include an inner diameter of 88 nm, an outer diameter of 104 nm, and a 30 nm gold thickness. Spatial distribution of the normalized magnetic field along Z generated by IFE in an XY plane 1 nm above the nanoring for b) right and c) left circularly polarized optical excitation, normalized by the **B** field amplitude at the center. d) and e) 3D vectorial distribution of unit vector **B** represented in b) and c), respectively. The black circular arrow at the right top of the figures indicates the polarization direction of the incident wave. f) Line cut of vectorial distributions of magnetic fields in amplitudes (B) and normalized (U) from the center to the edge of the nanostructure for right-handed (RH) and left-handed (LH) circular polarizations.

In conclusion, our theoretical investigation has demonstrated that the manipulation of light, specifically its polarization in the near-field of a plasmonic nanostructure, allows for the generation of a vectorial magnetic field distribution exhibiting the topology of a Neel-type skyrmion with a skyrmion number of ±1. This novel observation arises from exploiting the dark (or anti-bonding) mode in a ring-shaped gold nanostructure, allowing the generation of two counter-propagating drift currents in the inner and outer segments of the ring. We established that employing this dark mode enables the generation of identical elliptical polarizations on each side of the metal, giving rise to the counter-propagating currents. Furthermore, we demonstrated that a diverse set of parameters can be employed to achieve this dark mode and, consequently, the vectorial distribution characteristic of a skyrmion.

Ultimately, this distribution was illustrated through a case study illustrating the reversal of the magnetic field orientation in the spatial vector distribution from the center to the end of the plasmonic structure. This behavior aligns well with the anticipated characteristics of this skyrmion symmetry. The findings presented in this study represent a significant advancement in the generation and manipulation of nanoscale magnetic field distributions by IFE. This breakthrough opens avenues for directly implementing skyrmionic topological structures in magnetic materials through an all-optical approach, potentially operating at ultrafast timescales. Consequently, this work holds promise for applications in various fields, including manipulating magnetic processes, ultrafast magnetic modulation, magnetic trapping, spin currents, and spin precession, with direct applications such as ultrafast data writing and processing.

**Acknowledgements**

This work is supported by the ERC grant FemtoMagnet (grant no. 101087709), the Agence Nationale de la Recherche (ANR-20-CE09-0031-01, ANR-22-CE09-0027-04 and ANR-23-ERCC-0005), the Institut de Physique du CNRS (Tremplin@INP 2020) and the China Scholarship Council.

Supplementary Information

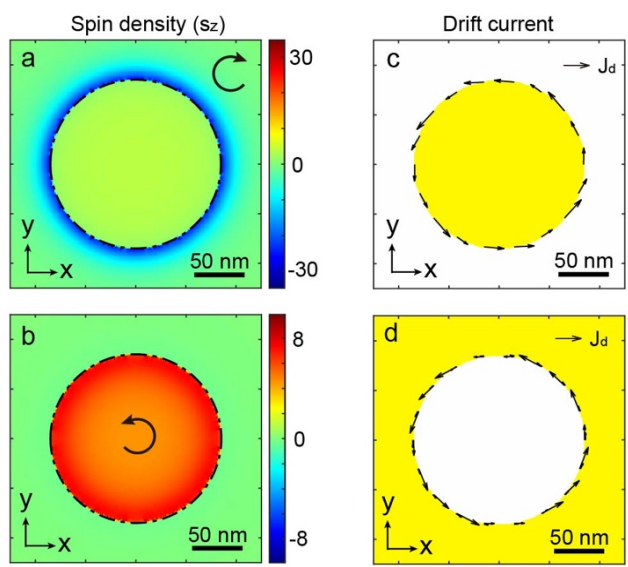

**Figure S1**. Comparison of optical properties of isolated nanostructures providing the nanoring: nanopatch and nanohole for a 30 nm gold thickness, deposited on a glass substrate, and for an excitation at λ = 754 nm and 825 nm with right circular polarization, respectively. Spatial distribution of spin densities in an XY plane at the center Z of a) a nanopatch and b) a nanohole. The arrows represent the helicity of light near the antennas. An important observation here is that, for the same excitation polarization of these plasmonic nanostructures, the local light polarization is opposite. c) and d) Spatial distribution of drift currents associated with the spin densities shown in a) and b), respectively. The length of the arrows represents the relative amplitude of drift currents.

Comprehensive information on the divergent polarization between the two bonding and anti-bonding modes.

As illustrated in Figures 2 and 4, the direction of the drift currents is associated with the z component of spin density. The spin density itself characterizes the handedness of the local elliptically polarized light, and the local polarization state, or the nearfield electric field, is directly influenced by the charge distribution in Figure 3c,d. Notably, the two intrinsic modes exhibit entirely different charge distributions.

In the case of the bonding mode, depicted in Figure S2a, it is composed of two parallel electric dipoles (ED). Each ED exhibits a spider-like electric field distribution in the near field.

Consequently, the total electric field results from the contribution of both EDs, producing the same electric field distribution as shown in Figure S2a. To assess the handedness of the nearfield, the orientation of the electric field at different times is indicated by black arrows in Figure S2a(ii, iv). Considering that the nano-ring is excited by right-handed circularly polarized (RHCP) light, the charge distribution and electric field undergo a 90-degree rotation after a quarter time period, as seen in Figure S2a(i,ii) and Figure S2a(iii,iv). In a complex representation, the electric field distribution in Figure S2a(i,ii) and Figure S2a(iii,iv) can be treated as the real and imaginary parts of the electric field. The change of orientation between Re(E) and Im(E) (alternatively between $t_0$ and $t_0+T/4$) reveals the handedness of local elliptical light, as depicted in Figure S2b. This distribution is further validated by numerical simulation results from Lumerical FDTD. Figure S3(a,b) and Figure S3(c,d) display the real (Re(E)) and imaginary (Im(E)) parts of the distribution at 1100 nm (bonding mode), respectively. The opposite handedness inside and outside the nano-ring aligns with the spin density distribution in Figure 4b.

A similar analysis is applied to the anti-bonding mode, but it is more intricate due to opposing EDs in Fig S2c. In this case, the total electric field combines contributions from the inner ($E_{inner}$ - red) and the outer ($E_{outer}$ - blue) ED. These two EDs exhibit destructive interference in the near field, and the orientation of the total electric field is determined by the relative strength of local $E_{inner}$ and $E_{outer}$. The strength of local $E_{inner}$ and $E_{outer}$ depends on the intensity of the ED and the distance to each ED source. According to the spectrum of charge density in Figure 3b, the anti-bonding mode appearing at 570 nm has a relatively stronger inner ED and a weaker outer ED. Consequently, $E_{inner}$ is stronger than $E_{outer}$ at positions farther from the two ED sources, which are the top, bottom, and middle positions in Figure 2c(i). Here, the solid line indicates a stronger local electric field, while the dashed line indicates a weaker local electric field. As for the left and right positions in Figure S2c(i), they are situated next to the outer ED. In this small region, $E_{outer}$ will be stronger than $E_{inner}$, benefiting from the shorter distance to the source. As a result, the total electric field has a nearly homogeneous distribution in the near field, as shown in Figure S2c(ii). Subsequently, the total electric field undergoes a 90-degree rotation after a quarter time period considering RHCP incident light, as depicted in Figure S2c(iii and iv). Figure S3(e,f) and Figure S3(g,h) display the real (Re(E)) and imaginary (Im(E)) parts of the distribution at 570 nm (anti-bonding mode), respectively. The homogeneous handedness inside and outside the nano-ring aligns with the spin density distribution in Figure 4a

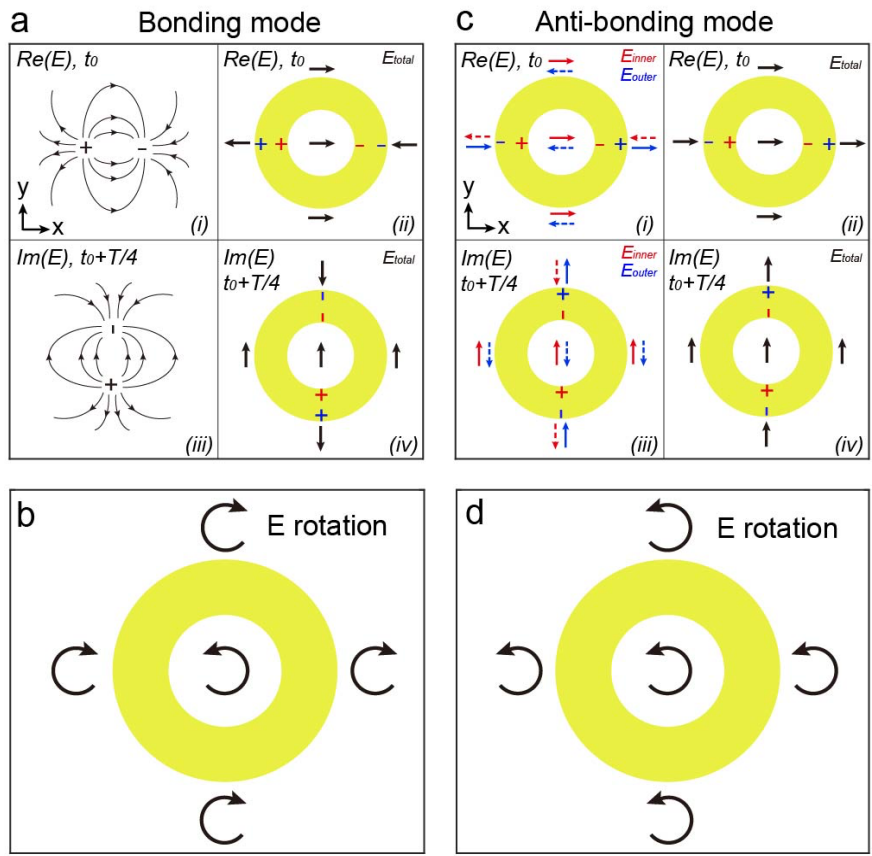

**Figure S2**. Dipolar study of bonding and anti-bonding modes. a) Spatial orientation in an XY plane of the electric field for i) a dipole oriented along X, ii) the nanoring excited by right circular polarization at $t_0$ and at the wavelength of the bonding mode, iii) an electric dipole oriented along Y, and iv) the nanoring excited by right circular polarization at $t_0 + T/4$ and at the wavelength of the bonding mode. b) Local polarization resulting from excitation of the nanoring by right circular polarization for the bonding mode. The arrows represent the helicity of light. c) Spatial orientation of the electric field resulting from the coupling between two opposite dipolar modes for excitation by right circular polarization of the nanoring at the wavelength of the anti-bonding mode at i, ii) $t_0$ and iii, iv) $t_0 + T/4$. In i) and iii), the red arrows represent the contribution of the inner dipole of the nanoring, the blue arrows represent that of the outer dipole, and the solid arrows represent the main contributions. In ii) and iv), the black arrows represent the orientation of the total electric field once the contributions of each dipole are taken into account. d) Local polarization resulting from excitation of the nanoring by right circular polarization for the anti-bonding mode. The arrows represent the helicity of light.

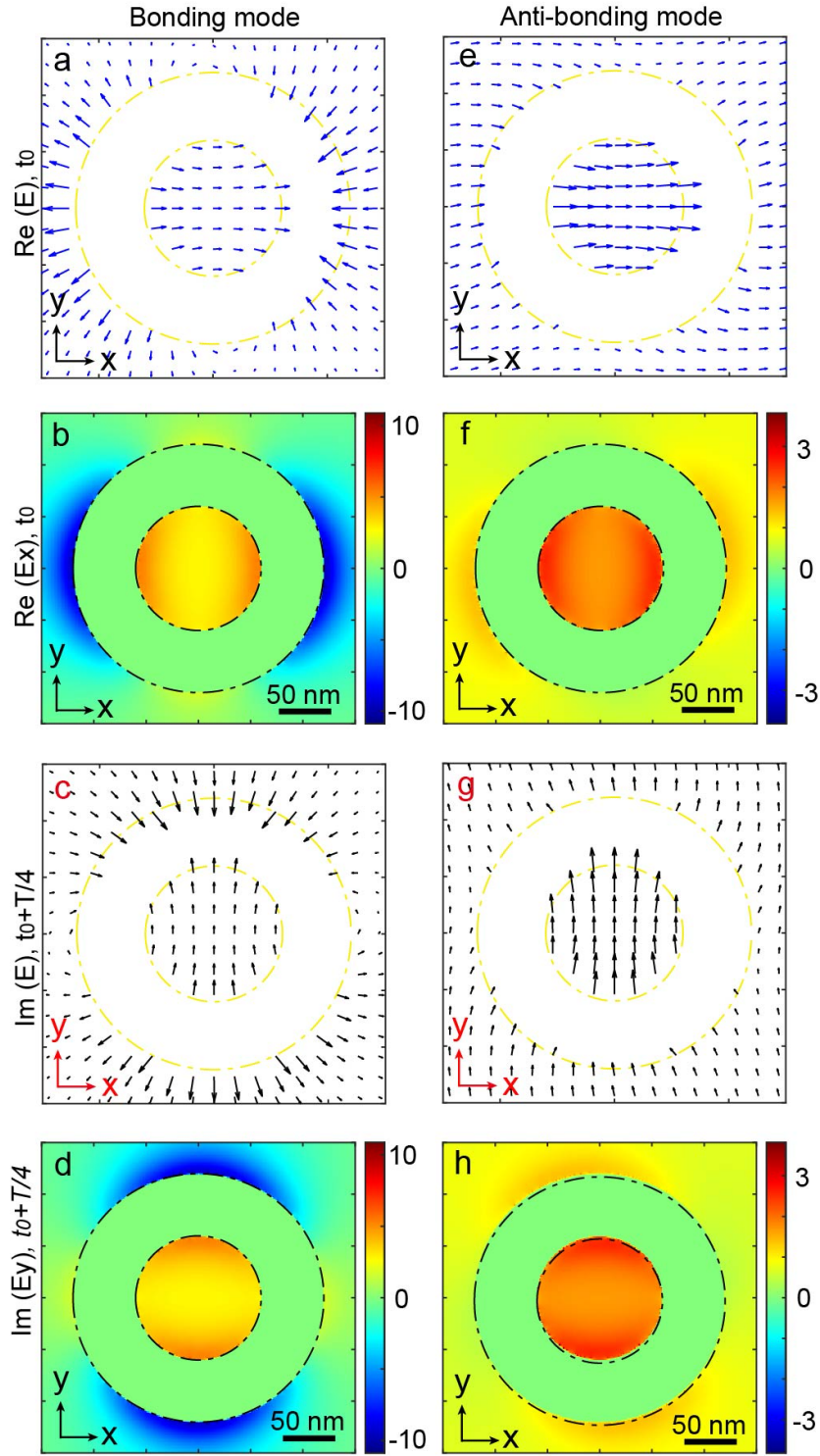

**Figure S3**. Distribution of electric fields in the nanoring for bonding ($\lambda$ = 1100 nm) and anti-bonding ($\lambda$ = 570 nm) modes at different times in an optical cycle and excited by right circularly polarized light. a, c) Vectorial distribution and b, d) electric field components in an XY plane at the center Z of the nanoring at a, b) $t=t_0$ and c, d) $t=t_0 + T/4$ for the bonding mode. e, g) Vectorial distribution and f, h) electric field components in an XY plane at the center Z of the antenna for times e, f) $t = t_0$ and g, h) $t=t_0 + T/4$ in the case of the anti-bonding mode. The length of the arrows represents the relative amplitude of electric field.

19